\documentclass[conference]{IEEEtran}
\IEEEoverridecommandlockouts
\usepackage{cite}
\usepackage{amsmath,amssymb,amsfonts}
\usepackage{algorithmic}
\usepackage{graphicx}
\usepackage{textcomp}
\usepackage{xcolor}
\def\BibTeX{{\rm B\kern-.05em{\sc i\kern-.025em b}\kern-.08em
    T\kern-.1667em\lower.7ex\hbox{E}\kern-.125emX}}
\begin{document}

\title{Conference Paper Title}
\title{Implant-to-Wearable Communication through the Human Body: Exploring the Effects of Encapsulated Capacitive and Galvanic Transmitters
%{\footnotesize \textsuperscript{*}Note: Sub-titles are not captured in Xplore and should not be used}https://www.overleaf.com/project/658dad17ddee0a7cab2882df
\\[-0.5ex]
}

% \author{\IEEEauthorblockN{Asif Iftekhar Omi}
% \IEEEauthorblockA{\textit{Dept. of ECE} \\
% \textit{University of Florida}\\
% Gainesville, USA \\
% as.omi@ufl.edu}
% \and
% \IEEEauthorblockN{Baibhab Chatterjee}
% \IEEEauthorblockA{\textit{Dept. of ECE} \\
% \textit{University of Florida}\\
% Gainesville, USA \\
% chatterjee.b@ufl.edu}}
\author{\IEEEauthorblockN{Anyu Jiang$^{1}$, Cassandra Acebal$^{2}$, Brook Heyd$^{1}$, Trustin White$^{1}$, Gurleen Kainth$^{2}$, Arunashish Datta$^{3}$, \\ Shreyas Sen$^{3}$, Adam Khalifa$^{1}$ and Baibhab Chatterjee$^{1}$}
\IEEEauthorblockA{$^{1}$\textit{Dept. of ECE,}, \textit{University of Florida,} Gainesville, USA. $^{2}$\textit{Dept. of BME}, \textit{University of Florida,} Gainesville, USA\\
$^{3}$\textit{School of ECE, Purdue University,} West Lafayette, USA\\
email: \{anyu.jiang, chatterjee.b\}@ufl.edu}

\\[-5ex]
}

\maketitle

\begin{abstract}
%Implanted devices that communicate with wearable monitors using traditional wireless telemetry suffer from high power consumption, high tissue absorption, and potential security issues.
%Conversely,
Data transfer using human-body communication (HBC) represents an actively explored alternative solution to address the challenges  related to energy-efficiency, tissue absorption, and security of conventional wireless. Although the use of HBC for wearable-to-wearable communication has been well-explored, different configurations for the transmitter (Tx) and receiver (Rx) for implant-to-wearable HBC needs further studies.
This paper substantiates the hypothesis that a fully implanted galvanic Tx is more efficient than a capacitive Tx for interaction with a wearable Rx. Given the practical limitations of implanting an ideal capacitive device,  we choose a galvanic device with one electrode encapsulated to model the capacitive scenario. We analyze the lumped circuit model for in-body to out-of-body communication, and perform Circuit-based as well as Finite Element Method (FEM) simulations to explore how the encapsulation thickness affects the received signal levels. We demonstrate in-vivo experimental results on live Sprague Dawley rats to validate the hypothesis, and show that compared to the galvanic Tx, the channel loss will be $\approx$ 20 dB higher with each additional mm thickness of capacitive encapsulation, eventually going below the noise floor for ideal capacitive Tx. %($\textgreater$ 2 mm encapsulation).

%With thick encapsulation ($\textgreater$ 2 mm), the Rx signal on the rat's back will below the noise level, incurring a $\textgreater$ 80 dB channel loss.
\end{abstract}


\vspace{-4mm}
\section{Introduction}
Implanted devices, such as smart glucose monitors and pacemakers, play an important role in managing certain health conditions
%detecting and treating diseases that are difficult to manage with conventional therapies.
in an effective manner. These devices can form a closed-loop system with an external %monitoring equipment, ideally a
smart hub on the human body, enabling the detection of specific health indicators. This integration leads to personalized therapies %or
with continuous monitoring, offering significant advancements in patient care. 

Currently, the wireless communication between implantable Tx and external Rx mostly relies on radio frequency (RF) protocols such as Bluetooth or Wi-Fi.
%Using such of mechanism will let the signal leaks outside from human body, having the risk of intercepting by hackers. Also, 
Because such traditional wireless techniques utilize high-frequency signals through a lossy medium (air),
%communicating under such a high frequency is power-hungry,
the energy-efficiency is usually poor ($\approx$ 1 nJ/bit \cite{chatterjee_arbio}), leading to frequent replacement of any battery in the Tx, or the need for better harvesting techniques. Additionally, traditional wireless gets absorbed in the body, and has a significant amount of electromagnetic (EM) leakage outside the body, making it susceptible to hacking \cite{das_scirep}.  HBC has emerged as a potential solution to these challenges, utilizing the body's conductive properties for secure, low-power data transmission \cite{zimmerman, wegmueller}. At Electro-quasistatic (EQS) frequencies ($\textless$ 10’s of MHz), HBC offers low-power and secure communication between devices in, on or around the body.

%Add some other people's work here - should we do this later, before talking about our contribution?

\begin{figure}[tbp]
    \centering
        \includegraphics[width=0.8\linewidth]{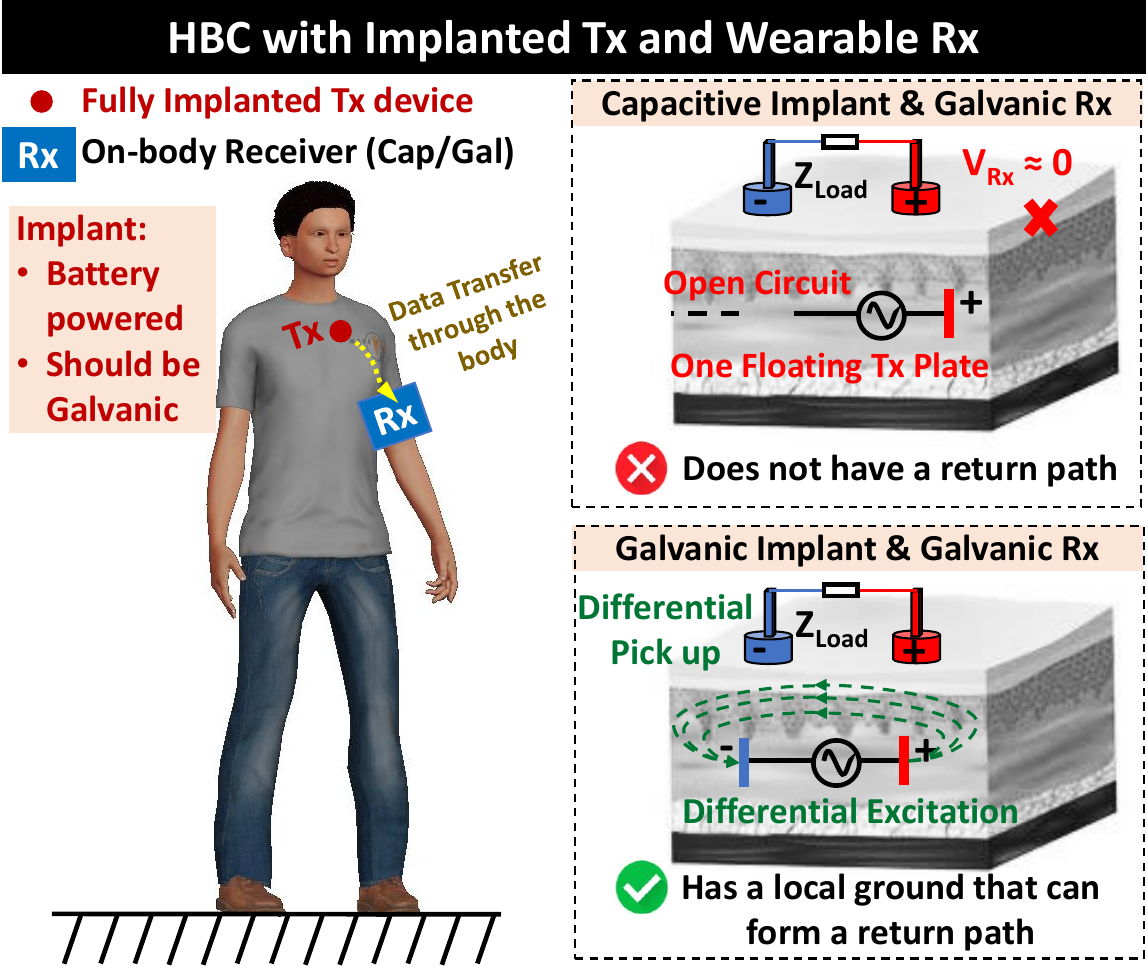}
        \vspace{-3mm}
        \caption{HBC for implanted Tx (capacitive/galvanic). Ideal capacitive Tx has no signal return path, which would result in almost no signal at the Rx. Note that only a galvanic Rx is shown (Rx can be either capacitive or galvanic).}
        \label{fig1}
        \vspace{-5mm}
\end{figure}

HBC technology, based on its %transfer and reception
signal coupling mechanisms, can be classified into two primary categories: capacitive\cite{zimmerman} and galvanic\cite{wegmueller}. 
Capacitive HBC operates through single-electrode coupling at the Tx and Rx, and transfers electrical signals through the forward path within the human body. The return path is formed by the parasitic capacitances between the earth ground and the local reference/ground planes in the Tx and Rx, as well as the parasitic capacitance formed between the human body and the earth ground \cite{maity_TBME_2019}. 
%Meanwhile has has a parasitic capacitor that exists between the earth's ground and the floating ground plate of the device, that forms the forward path and the return path of the Tx and Rx. 
For galvanic HBC, on the other hand, the Tx and Rx are differential (coupled to the body using two electrodes/dipole coupling). The electric fields generated in the Tx flow through the body and are subsequently captured by the differential receiving device. 

For wearable applications, capacitive HBC is a safe and simple approach. However, importantly, for a fully implanted HBC device, %a capacitive Tx will never work.
a capacitive Tx will incur a significantly higher amount of loss as the local ground plane at the Tx will not have a direct parasitic return path to the earth ground \cite{Datta_IMS21, Johnson_EMBC23}.

\begin{figure}[htbp]
    \centering
        \includegraphics[width=0.8\linewidth] {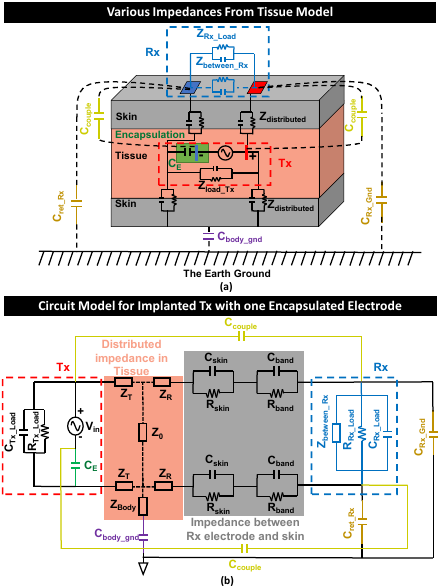}
        \vspace{-5mm}
        \caption{(a) Tissue and skin impedance model with implanted Tx (with one encapsulated electrode) and galvanic Rx. %Explaining various of impedance, like the impedance inside the body and the impedance from outside.
        (b) Circuit model of (a). Note that the impedances inside the body are distributed in reality.}
        \label{fig2}
        \vspace{-6mm}
\end{figure}

As shown in Fig.\ref{fig1}, a capacitive fully implanted Tx does not form a close loop signal path with the Rx, leading to the Rx voltage $\approx$ 0. Consequently, a fully implanted Tx needs to be galvanic. In practical scenarios, implanting an ideal capacitive Tx with a floating reference electrode that doesn't interact with the surrounding tissue is hard to achieve. One approach to model the capacitive scenario is to encapsulate the reference electrode of a galvanic Tx. Depending on the encapsulation material and thickness, the reference electrode on the Tx %will stimulate
can still excite the tissue through capacitive dipole coupling. However, if the encapsulation is thick enough, the Tx can be considered as capacitive.

An extremely important consideration for measuring in-vivo HBC performance for implants is that the setup should have a fully implanted Tx with closed surgical wounds, which would ensure that no component of the Tx (not even the battery or any wires) is outside the body, which would otherwise capacitively couple with the Rx, and offer optimistic results even in the capacitive Tx scenarios through additional coupling \cite{Bae_SSCL20, Chae_JSSC22}. Additionally, encapsulating the reference electrode in the Tx may not ensure capacitive HBC operation, as galvanic dipole coupling will still be dominant with some additional loss due to the encapsulation, very similar to the biphasic coupling scenarios \cite{Johnson_EMBC23, chatterjee2023biphasic}.

%\textcolor{red}{Need to nicely talk about a few works that have shown capacitive transmitters in the past.}

In this paper, we explore the channel transfer function (TF) for both a galvanic and capacitive (galvanic with the reference electrode encapsulated) Tx, along with supporting results from an equivalent circuit model, FEM simulations as well as in-vivo experiments. % about why a capacitive implant will not work practically is analyzed based on the encapsulated Tx.
The contributions of this study are highlighted as follows:
\begin{itemize}
\item Developed a \textbf{circuit model} for fully implanted Tx and wearable Rx, based on the impedances present for in-body to out-of-body signal transfer, which explains how the encapsulation thickness for the reference electrode on the Tx affects the channel loss. Circuit simulations in Cadence show why an ideal capacitive Tx is unsuitable.
\item Performed \textbf{FEM simulations} for the implant in a simplified rat EM model with multiple tisue layers using Ansys High Frequency Structure Simulator (HFSS).
\item Finally, developed a fully implanted, flexible signal generator as an HBC Tx and performed \textbf{in-vivo experiments} with galvanic as well as encapsulated Tx in live Sprague Dawley rat models to observe channel losses. This crucial measurement confirms that the theory holds true in a living organism. The fully implanted Tx helps in avoiding any parasitic capacitive effects during signal transfer. 
\end{itemize}

The rest of the paper is organized as following: Section II presents a critical analysis of the circuit model and FEM simulation results for capacitive (encapsulated) vs galvanic HBC. Section III details the setup and results of the in-vivo measurements. Finally, Section IV concludes the paper.

\section{Theory, Analysis and Simulations}
%The focus of this work is to prove
An implanted capacitive Tx (with one encapsulated and one unencapsulated electrode) will suffer from a significantly high loss at EQS frequencies (10s' of MHz or lower). %Practically, a implanted capacitive Tx is achieved by encapsulating one electrode of galvanic Tx.
In this section, the circuit model for such scenarios is analyzed based on implanted encapsulated Tx and wearable galvanic Rx. The modality of the wearable Rx will not have a strong impact on the %function of sensing the power
working principle of the Tx \cite{Datta_IMS21}. The effect of encapsulation can be observed through %the simulation of proposed circuit model and EM model of the rat.
circuit-level simulation of an electrical model in Cadence as well as through FEM simulation of a rat model in HFSS.

\vspace{-0.5mm}
\subsection{Implanted Tx Circuit Model}
\vspace{-1mm}
There have been recent studies\cite{maity_TBME_2019,Modak_TBME22} to characterize the circuit model of both capacitive and galvanic Tx/Rx in wearable scenarios.
Such circuit models utilize impedances based on the design of the Tx and Rx, as well as tissue impedances and parasitic impedances (return paths to earth ground for capacitive HBC), the values for which are calculated based on %electronic constant of 
the permittivity and conductivity of the tissue and surrounding media \cite{S_Gabriel_1996}, along with real measurements. %However, as for galvanic Tx inside the tissue, there is no direct coupling from the Tx patch to the earth ground.
For an implanted Tx, the circuit models needed to be updated. Also, previous models are based on the tissue dielectric properties at 100's of kHz, which means that the value for each component needs to be scaled when it is operated at any other frequency, such as $\sim$21MHz (IEEE 802.15.6 standard for body area networks).

% \begin{figure*}[htbp]
%     \centering
%         \includegraphics[width=0.9\linewidth]{Fig4+3.jpg}
%         \caption{(a) Top view and (b) side view of simplified rat model, showing the implement of Tx and Rx. (c) Pure galvanic Tx and encapsulated Tx to mimic the capacitive implant Tx. (d) Modality of galvanic Rx. (e)Result of the simulation for the encapsulated Tx with both HFSS and Cadence. Note that X axis stands for the increasing of the encapsulation thickness or the deduction of the $C_{E}$.}
%         \label{fig3}
%         \vspace{-5mm}
% \end{figure*}

The different impedances in the circuit model is shown in Fig.\ref{fig2}(a). %The Tx transmits the signal by connecting two electrodes that directly excite the tissue. In between, there will be tissue act as parallel RC loading the source, which will occupy fractions of the power transferring\cite{chatterjee2021biphasic}.
% we are not doing power transfer
The Tx directly couples signal to the tissue using two electrodes in the galvanic scenario, while the encapsulation in capacitive scenario is modeled by the capacitance, $C_E$ on the reference electrode of the Tx. The Tx is loaded by parallel tissue impedances, which are modeled by the parallel RC to the Tx. 
As shown in Fig.\ref{fig2}(b), %From the Tx electrode to the skin layer under the Rx electrode,
a distributed tissue impedance model is considered from the Tx electrodes to the skin layer under the Rx electrode. In reality, the impedance value will vary with the distance from Tx to Rx. Because the capacitance in between is low, the tissue of the forward path can be simplified as resistance dominated. %pure resistance
%For the gap between the tissue and the skin and the air gap between Rx electrodes and the outer skin, both of them 
The skin layer and the interface of the Rx electrodes with the skin are represented as parallel RC. Additionally, the inherent parallel RC load at the Rx device, along with a parallel resistance between the 2 Rx electrodes (representing the conductive path through the skin) will form the total Rx load, note that the resistance between can be adjusted by changing the distance.

%%%%%%%%%%%%%%%%%%%%%%%%%%%%%%%%%%
%Baibhab will check the rest on Thursday

 %ensure the circuit form a close loop
 
 % very imp: Please note that the parasitic caps create additional coupling from Tx to Rx, but not required to create the closed loop path as both your Tx and Rx in the model are galvanic
% To establish a common ground, the entire system requires a return path, 
%The path brings the electrode to the common ground is formed by the coupling capacitance between the electrode and the earth. %to the earth ground. 
%At the Tx side, the Tx electrodes have a high resistance path ($Z_{body}$) from them to the skin layer of the body, then the outer body will have a coupling capacitor to the earth ground. 
%For the Rx, the total capacitance connected to one electrode towards the ground comprises two paralleled components: the intrinsic capacitance of the electrode to the ground($C_{ret\_Rx}$), and the capacitance between the body and the ground($C_{Rx\_Gnd}$).
In addition to the differentially coupled signal path between the Tx and Rx, there will be parasitic capacitance from the wearable Rx electrodes to the earth's ground ($C_{ret\_Rx}$ and $C_{Rx\_ Gnd}$, as shown in \cite{maity_TBME_2019}), as well as parasitic capacitances representing inter-device coupling $C_{Couple}$ between the Tx and Rx. These parasitic capacitances introduce additional coupling from Tx to Rx. However, unlike the capacitive wearable HBC case, they are not necessary for forming any closed-loop path for signal transfer, when both the Tx and the Rx are galvanic.
%These parasitic capacitors, however, unlike a capacitive wearable case, are not required to forming a closed loop path to enable signal tranfer. Since both the Tx and the Rx are considered galvanic. 

%As a result, the entire system has a common earth ground, establishing a closed-loop signal path that enables the passage of the signal.
%For the Rx, the total capacitance attach to one electrode to the ground is the capacitor of the electrode itself to the ground $C_{ret_Rx}$ parallel with the capacitance from the body to the ground $C_{Rx_Gnd}$. 

%One of the electrodes of the galvanic Tx is encapsulated in order to act like a capacitive Tx.%not directly excite the tissue. %As the encapsulation getting thicker, it will more like an ideal capacitive Tx. 

To model the encapsulation on the reference electrode of the Tx, a series capacitor ($C_{E}$) is introduced at the local ground of the Tx. %This capacitor is formed between that electrode and  tissue, with the encapsulation serve as dielectric.
In theory, the thicker the encapsulation, the smaller will be the value of $C_{E}$ (larger impedance), leading to additional loss. Considering the electrode and the encapsulation as two concentric cylinders, the capacitance of this configuration can be computed, typically within the pF range.
%Assuming that the electrode and the encapsulation are two concentric cylinders, that capacitor can be calculated to be at the magnitude of pF. 

%With increased encapsulation thickness, the distance between the electrode and the tissue enlarges, resulting in a reduced capacitance at the interface.
In the limiting scenario, $C_{E}=0$ (infinite encapsulation) behaves like an open circuit% from the refrence electrode to the tissue, at frequencies in the order of 10s' of MHz
. As a result of such single-electrode coupling, the Tx can be considered as capacitive, leading to very weak Rx signals, as the entire voltage drop happens across the impedance due to $C_{E}$. 

%In addition to HBC, the Tx and Rx also engage in communication through inter-device coupling. In the circuit model, this is represented by a capacitance ($C_{Couple}$) directly connecting the Tx and Rx nodes. Inter-device coupling plays a crucial role in limiting the maximum loss of the channel, thereby influencing the overall system performance.

%As the encapsulation is very thick, the distance between and the electrode is large, thereby there will be a very small capacitance in between the tissue and the electrode, act like a open circuit under 10's of MHz of frequency. As a result, the model can be consider as capacitive Tx, lead to very small signal received. 

\vspace{-0.5mm}
\subsection{Simulation Setup and Result}
\vspace{-1mm}
To validate the effect of encapsulating one electrode in the Tx, a circuit-EM co-simulation for channel loss was performed. This involved a lumped circuit model simulation using Cadence, coupled with FEM simulations in Ansys HFSS.

%To validate the thickness of encapsulation on the galvanic implant device will affect the channel loss, co-simulation with lumped element model using Cadence as well as FEM based electro-magnetic simulation on HFSS is performed.
%The received voltage along the back of the rat, with galvanic Rx and galvanic Tx with/without encapsulation will be simulated as well. % talk about the distance between the Rx and Tx
\subsubsection{Simulation Setup}
The lumped element circuit model of encapsulated Tx and galvanic Rx is shown in Fig.\ref{fig2}(b). To analyze loss at frequencies $\approx$ 21 MHz, component values are estimated based on the Gabriel model \cite{S_Gabriel_1996} and \cite{maity_TBME_2019}. To align the encapsulation thickness effect to the Rx signal, the value of $C_{E}$ was adjusted within pF range. %By applying an 21 MHz AC signal with magnitude of 1 V, the channel loss at the Rx side can be quantified. 
The channel loss was measured near 21 MHz according to the IEEE 802.15.6 standard.

\begin{figure}[htbp]
    \centering
        \vspace{-2mm}
        \includegraphics[width=0.9\linewidth]{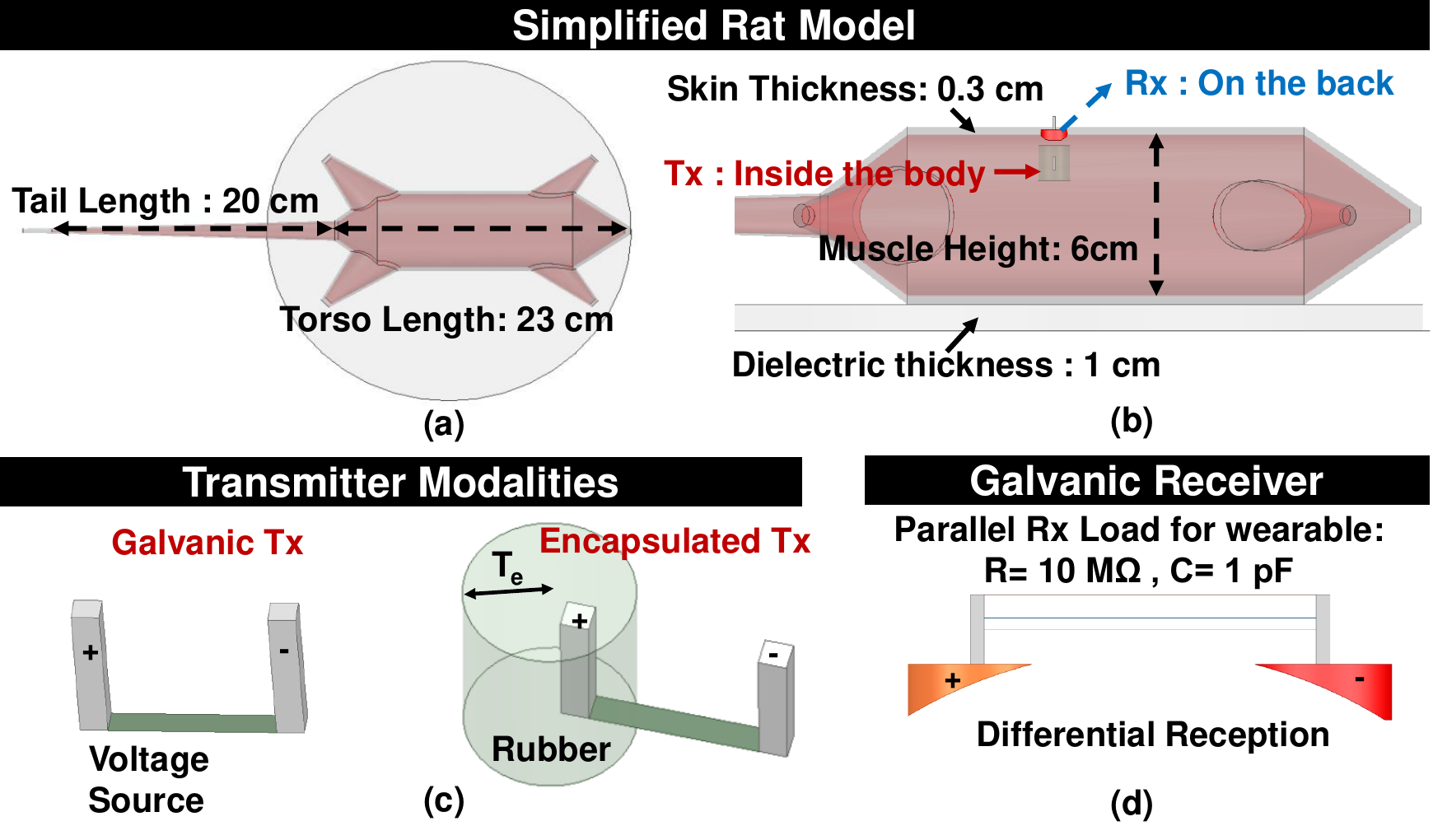}
        \vspace{-4mm}
        \caption{(a) Top view and (b) side view of the simplified rat model, showing the implementation of Tx and Rx. (c) Pure galvanic Tx and encapsulated Tx to mimic the capacitive implant Tx. (d) The Galvanic Rx model. }
        \label{fig3}
        \vspace{-2mm}
\end{figure}

Also, a simplified model of a rat made of skin and muscle tissue is created in HFSS for the EM simulation, which is shown in Fig.\ref{fig3}(a)-(b). First, a galvanic Tx is placed inside the rat's body while a galvanic Rx is place on the skin of the rat. The Tx %stimulates the body using two electrodes, 
excites the tissue differentially with a 1 V lumped voltage source applied. The Rx voltage is calculated by integrating the electric field along an integration line between the Rx electrodes.
%The Tx excites the body with two thin electrodes with a voltage source of 1 V connected in between. 
Next, a hard rubber ($\epsilon$ = 3) is attached around the Tx reference electrode to make the Tx act capacitively. The encapsulation thickness will be adjusted to observe its corresponding effect on channel loss, which is the ratio of Rx voltage to Tx voltage for voltage-mode HBC.
%The thickness of encapsulation will be adjusted to observed how will that affect the channel loss. 

\begin{figure}[htbp]
    \centering
        \vspace{-2mm}
        \includegraphics[width=0.7\linewidth]{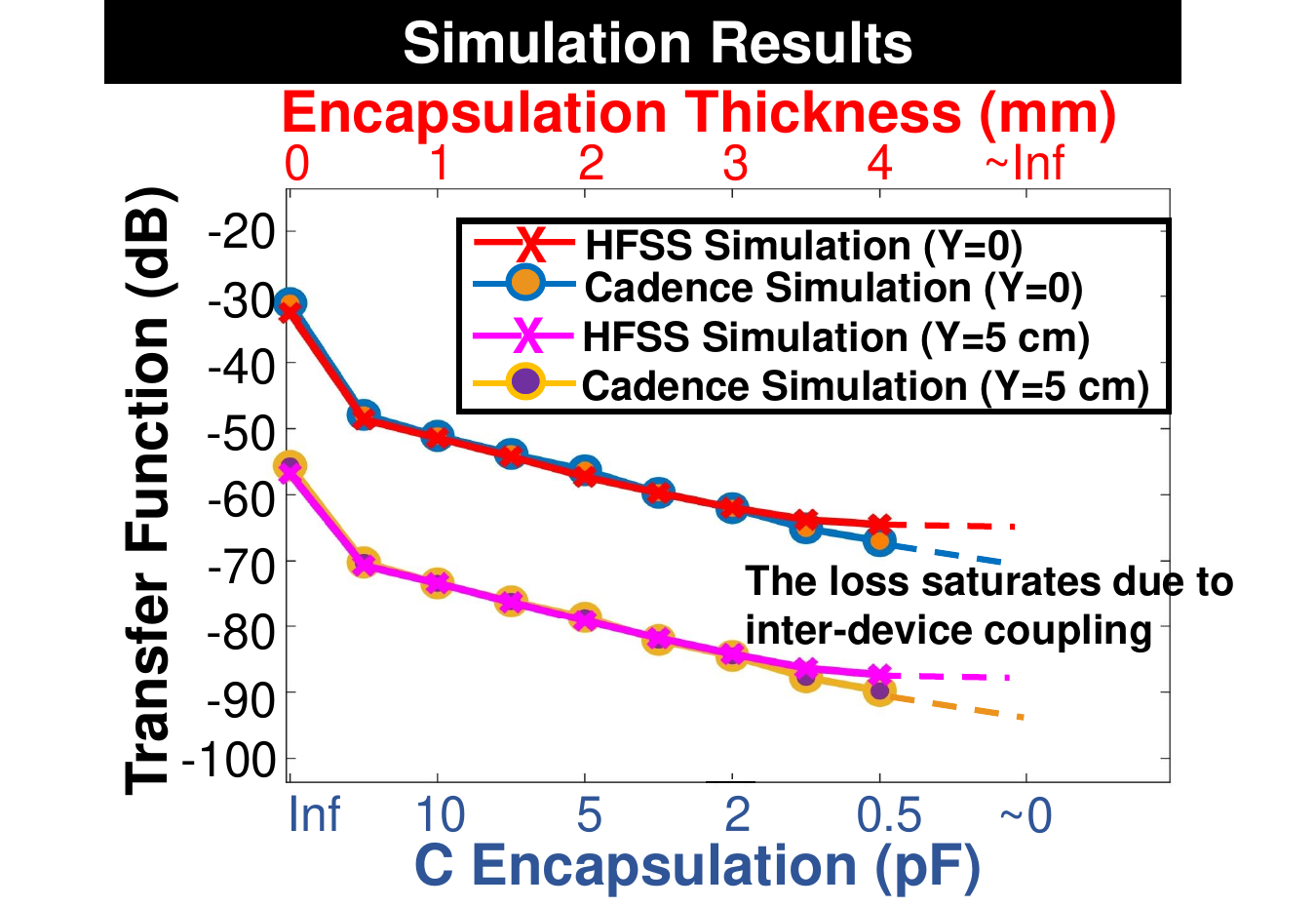}
        \vspace{-3mm}
        \caption{Simulation Result for the encapsulated Tx with both HFSS and Cadence. The x-axis is the encapsulation thickness (or the value of $C_{E}$). %Note that the Y-axis position of the Tx inside the rat's body is set at 0.
        }
        \label{fig4}
        \vspace{-2mm}
\end{figure}

\subsubsection{Simulation Result}
The simulation results, as illustrated in Fig.\ref{fig4}, explain the correlation between the channel loss and either the value of $C_{E}$ or the thickness of encapsulation. Key observations include the following: Initially when the encapsulation is thin ($C_{E}$ is large), %the Tx still works with some addition loss. 
a substantial amount of signal can still be received through dipole coupling. %Because the encapsulated electrode will work under dipole coupling, where  value. Further,
As the encapsulation gets thicker, $C_{E}$ decreases, introducing greater impedance as well as signal loss at the encapsulation, causing the Rx voltage to drop. In the extreme scenario of very thick encapsulation, the %dipole % one electrode cannot have dipole coupling
coupling between the encapsulated electrode and the tissue becomes negligible, indicating that $C_{E}$ $\approx$ 0. Hence, the Tx can be considered as capacitive, leading to its inefficacy for implantation purposes. Additionally, depending on the distance between the Rx and Tx, still a considerable amount of signal can be detected due to inter-device coupling, %even the HBC effect is minimized. Inter-device coupling leads to
which sets a saturation limit in the channel loss even when $C_{E}$ $\approx$ 0, or equivalently, when the encapsulation becomes significantly thick, as seen from the simulations.

% \begin{figure*}[htbp]
%     \centering
%         \includegraphics[width=\linewidth]{fig5+6.jpg}
%         \vspace{-5mm}
%         \caption{(a) RF Explorer handheld spectrum analyzer with buffer, with 2 jumper wires uses for connecting to multi-electrode Rx. (b) The structure of flexible Tx implant, it is a battery powered ring oscillator. (c) Side view with the set up of galvanic Tx and encapsulated Tx.(d) Multi electrode array that will be put on the rat's back to sense the power transmitting. (e)The diagram of the experiment system. (f) The experiment set up of the in vivo measurement on rat.(g)3D plot for Tx with different thickness of encapsulation versus(a) Capacitive Rx, (b) galvanic Rx.Note that if the encapsulation is greater than 2 mm, the Rx signal will below the noise level and the channel loss will greater than 85 dB.}
%         \label{fig5}
%         \vspace{-7mm}
% \end{figure*}

% \section{Experimental Results}
\section{In-vivo Experiment}
The in-vivo experiments were conducted using
anesthetized Sprague Dawley rats over multiple days according to proper Institutional Animal Care and Use Committee (IACUC) guidelines at University of Florida, and repeatable results are reported.
The experiments confirm that there is a corresponding increase in channel loss as the encapsulation on one electrode of the galvanic Tx gets thicker. This hypothesis holds true for both capacitive and galvanic Rx on the rat's body.

\subsubsection{Experiment Setup}

\begin{figure}[tbp]
    \centering
        \vspace{-2mm}
        \includegraphics[width=\linewidth]{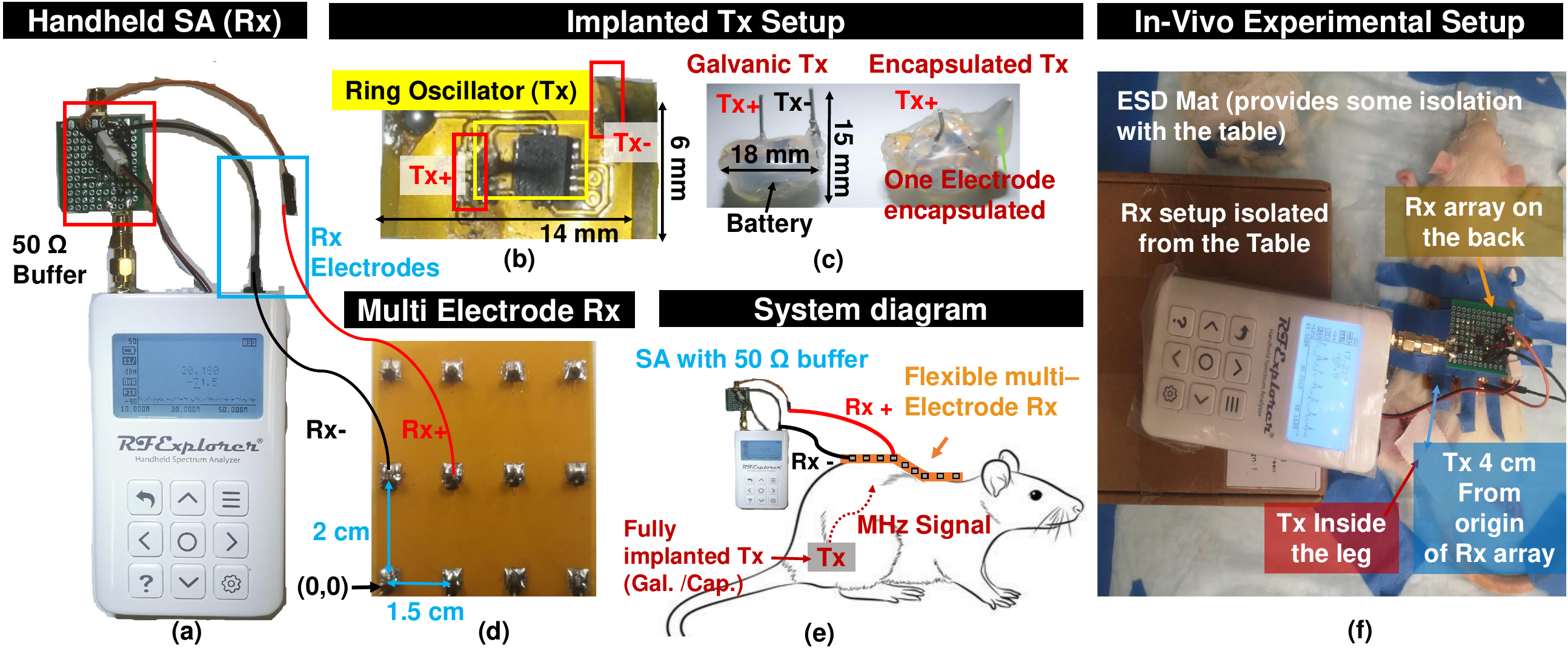}
        \vspace{-5mm}
        \caption{(a) RF Explorer handheld spectrum analyzer (SA) with 50 $\Omega$ buffer, connecting to a multi-electrode Rx. (b) The Flexible Tx Implant, featuring a battery-powered ring oscillator. (c) Side view of the set up, showing galvanic Tx and encapsulated Tx. (d) Multi electrode array taped on the rat's back to measure Rx signal. %(a) RF Explorer handheld spectrum analyzer (SA) with 50 $\Omega$ buffer, connecting to multi-electrode Rx. (b) Structure of the Flexible Tx Implant, featuring a battery-powered ring oscillator. (c) Side view of the set up, showing galvanic Tx and encapsulated Tx.(d) Multi electrode array that will be taped on the rat's back to sense the power transmitting.
        (e) Full system diagram. (f) In-vivo measurement setup.}
        \label{fig5}
        \vspace{-5mm}
\end{figure}

The % whole
in-vivo experimental set up is detailed in Fig.\ref{fig5}. The Rx consists of a handheld spectrum analyzer (SA) connecting to a 50 $\Omega$ buffer (Texas Instruments BUF602). This buffer drives the 50 $\Omega$ SA while offering high-impedance termination at the body \cite{Datta_IMS21}.
%The 50 $\Omega$ buffer is to match with the input impedance of the SA, getting rid of the issue that high impedance termination on the human body occupies most of the power.
A 3$\times$4 electrode array, made with flexible Polyimide PCB, is connected to the skin of the rat's back for robust and reconfigurable signal reception at various locations. Each electrode on the array is a square with 5mm$\times$5mm surface area.
%The array will connect to the SA using jumper wire, which makes the Rx system more robust and reconfigurable. 

The fully implanted galvanic Tx is designed on a flexible Polyimide PCB using the SN74AHC04BQAR inverter. By setting the Tx as a 3-stage ring oscillator, $\approx$ 21 MHz signal can be generated. The galvanic Tx has 2 exposed electrodes, while the remaining parts are sealed to ensure waterproofing. To mimic the conditions of a capacitive implant, the reference electrode is coated with glue to prevent direct contact with the rat's tissue. To ensure the accuracy of the results, both the rat and the Rx are isolated from the table, minimizing potential earth coupling \cite{chatterjee2023biphasic}.
%To mimic the capacitive implanted device, one of the electrodes is encapsulated with glue, making it not directly connecting with the tissue of the rat. Both the rat and the Rx are isolated from the table to achieve an accurate result.

%All research protocols were approved and monitored by the University of Florida Institutional Animal Care and Use Committee.
% I put this in the beginning
 
For the experiments, both female and male adult Sprague Dawley rats were used. Throughout the surgical process, the rats were anesthetized with 1\%–3\% isoflurane. The surgical procedure involved an incision that ran from the knee joint to the ischial tuberosity to expose the muscle of the leg of the rat. After that, the Tx will be placed under the skin, with the surgical wound closed, with one (capacitive) or two (galvanic) Tx electrodes making direct contact with the muscle tissue.
%The Tx was then put under the skin of leg, with electrode that exciting the muscle.
%The device was then attached to the sciatic nerve; during the procedures, the Tx remained inside the rodent while the MEA was moved on the back of the rodent to collect appropriate data. 

The experiment measures the channel loss on the back of the rat with a fully implanted Tx. The bottom left corner pin of the multi-electrode Rx was considered as the origin for the coordinate system, %Set the bottom left corner pin of the multi-electrode Rx as a origin point, 
the Rx can alternate between capacitive and galvanic modes by connecting either one or two electrodes. This setup allows for the observation and measurement of channel loss variations in response to changing the encapsulation thickness of the Tx as well as the modality of Rx.
At the experimental endpoint, each rodent was euthanized.

\subsubsection{Experimental Results}

\begin{figure}[tbp]
    \centering
        \vspace{-2mm}
        \includegraphics[width=\linewidth]{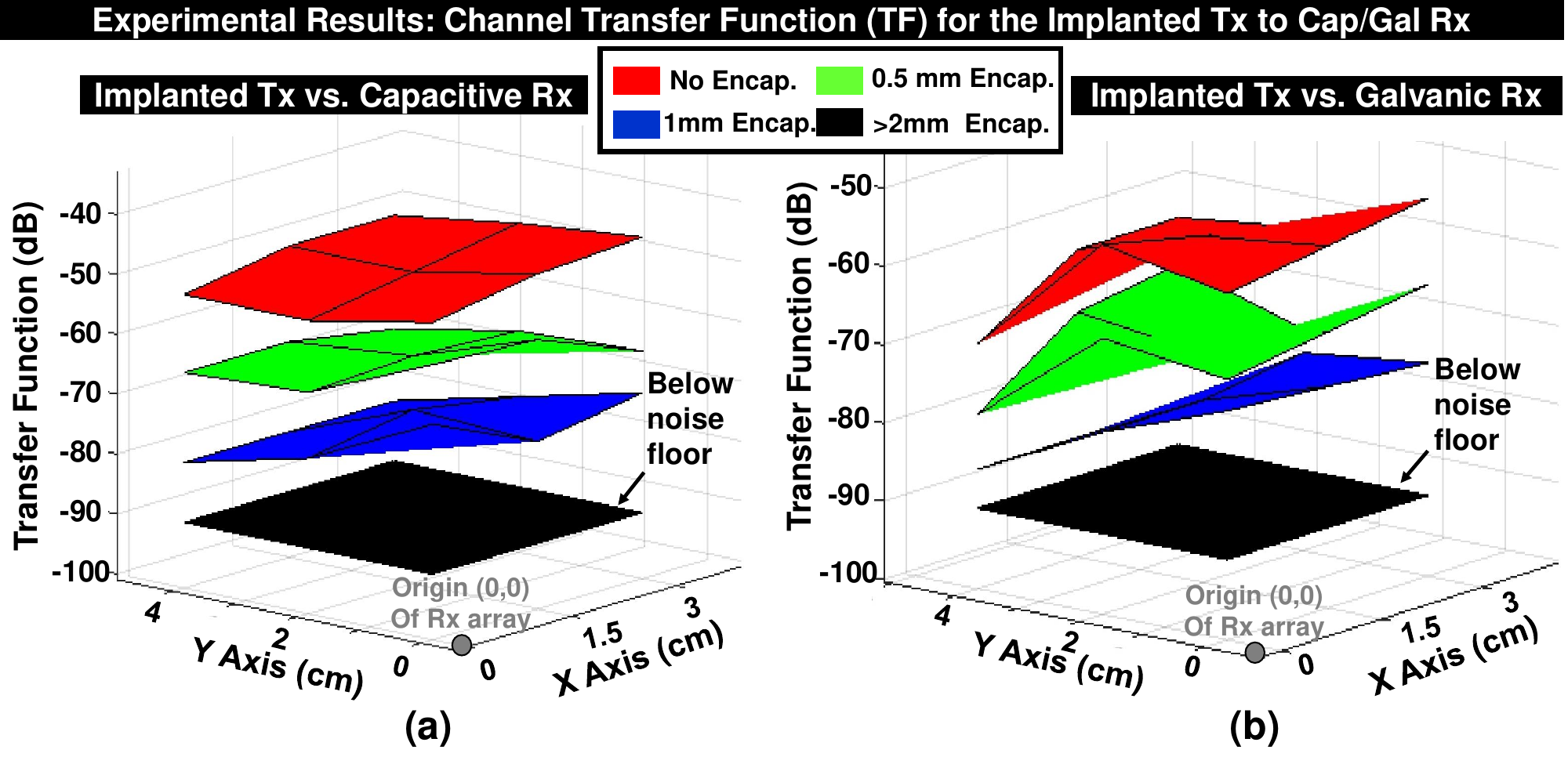}    
        \vspace{-5mm}
        \caption{3D plot for Tx with different thickness of encapsulation versus(a) Capacitive Rx, (b) Galvanic Rx with electrodes spaced 1.5 cm apart, moving along the back. Note that if the encapsulation is greater than 2 mm, the Rx signal will below the noise level and the channel loss will greater than 85 dB.}
        \label{fig6}
        \vspace{-5mm}
\end{figure}

The %experiment
experimental results are shown in Fig. \ref{fig6}. The channel loss measured along the back of the rat %through the Rx set up has been plotted 
is shown as a surface plot. For both capacitive and galvanic Rx, as the encapsulation gets thicker, the channel loss keeps increasing at a rate of $\approx$ 20dB per mm of encapsulation, eventually going below the SA noise level ($>$ 80 dB loss).
%Initially when the encapsulation is thin, the Rx signal got attenuated by 10 dB comparing to the galvanic Tx because the value of $C_{E}$ is still large.
As the encapsulation thickness increases, the implant begins to behave more like a capacitive Tx, and a high channel loss is observed. This trend is consistent with the data from the simulation shown in Fig.\ref{fig4} (although the slope is different because of slight material property mismatch), thereby validating the measurements obtained in the experiment. Also, the trend of the channel loss corresponding to the 2D distance from Tx with different modality of Rx matches with previous study \cite{Datta_IMS21} - with increasing distance, capacitive Rx exhibits a saturation in channel loss, whereas galvanic Rx shows a continuous reduction in the channel TF.

\section{Conclusion}
%In this project, 
We have investigated the effects of capacitive and galvanic Tx for implant-to-wearable scenarios in HBC using a proposed lumped circuit model, FEM simulations and in-vivo measurements, which demonstrate well-aligned results.
%For implant-to wearable HBC, the use of capacitive like Tx with a galvanic receiver has been analyzed through a proposed circuit model. To test our hypothesis concerning the efficacy of a capacitive implant, we conducted a co-simulation that integrated a lumped circuit model simulation with EM model simulations.  %A co-simulation with lumped circuit model and EM has been operated to validate the hypothesis for a capacitive implant.
The in-vivo experiment has been performed with both capacitive (encapsulated) and galvanic Tx and Rx. The experimental results indicate that with 1 mm encapsulation on one of the Tx electrodes, $\approx$ 20 dB loss will be added compare to a pure galvanic Tx. %Moreover, when the Tx is act as capacitive, which means the encapsulation has a substantial thickness, the Rx signal eventually falls below the noise level.
With $>$ 2mm encapsulation thickness, the Rx signal goes below the noise levels.
%And the Rx signal eventually goes below the noise level when the Tx is considered capcacitive. 
%We can prove the hypothesis that for implant-to-wearable scenarios, a galvanic Tx will be more effective, a fully implanted capacitive Tx lacks of feasibility for interaction with a wearable receiver.
This study substantiates the hypothesis that, in implant-to-wearable scenarios, a galvanic Tx proves more efficient, while a fully implanted capacitive Tx is not feasible for effective interaction with a wearable Rx.

\bibliographystyle{IEEEtran}
\bibliography{EMBC_2024_V1}

\end{document}